# Characterisation, Raman, Magnetic and Resistivity measurements in polycrystalline samples of LaMnO$_3$ doped with Cd

J. López[1,4], V. Dediu[1], P. Nozar[1], G. Ruani[1], C. Dionigi[1], F. C. Matacotta[1], C. Ferdeghini[2] and G. Calestani[3]

[1]ISM, CNR, Via Gobetti 101, 40129, Bologna, Italy
[2]Università di Genova, Dipartimento di Fisica, Via Dodecaneso 33, 16146 Genova, Italy
[3]Università di Parma, Dipartimento di Chimica, Viale delle Sciencie, 43100, Parma, Italy
[4]new address: D. Física, U. Federal de São Carlos, S.P, 13565-905, Brazil, e-mail: pjlopez@iris.ufscar.br

We report a study of polycrystalline samples of the family La$_{1-x}$Cd$_x$MnO$_{3+\delta}$ with different percentage of Mn$^{4+}$ ions. X-rays diffraction, Iodometric titration, Raman, Magnetic and Electrical Resistivity measurements provide a general characterisation of the physical properties. Results are qualitatively similar to the ones found in Ca doped manganese perovskites.

## 1- Introduction

The substitution of divalent ions of Ca, Sr, Ba, and Pb (position A') for the trivalent ions of La, Nd, Pr (position A) in compounds of the family A$_{1-x}$A'$_x$MnO$_3$ has been studied intensively during the last years[1-3]. An interesting feature of some of these materials is a paramagnetic insulator to ferromagnetic metal transition, accompanied by a high Magnetoresistance near the Curie temperature (Tc). Troyanchuk et al.[4,5] have reported structural, magnetic and resistivity measurements in La$_{1-x}$Cd$_x$MnO$_3$ compounds. They found a paramagnetic to ferromagnetic transition, but not an insulator metal transition, in samples with x between 0.25 and 0.5, and interpreted their results in the framework of a superexchange interaction. On the other hand, M. Sahana et al.[6] studied polycrystalline and thin film samples of La$_{0.67}$Cd$_{0.33}$MnO$_3$ and found an insulator metal transition at the same temperature interval of the paramagnetic ferromagnetic transition. As far as we know, these are the only reports available in the literature about Cd substitutions in manganese perovskites. It would be interesting to investigate further whether the doubleexchange or superexchange interaction is responsible for the physical properties in Cd doped manganese perovskites and make a comparison with closely related Ca doped compounds. Here, we report a study of four polycrystalline samples of the family La$_{1-x}$Cd$_x$MnO$_{3+\delta}$ with different percentage of Mn$^{4+}$ ions. We used X-rays diffraction (XRD), Iodometric titration, Raman, Magnetic and Electrical Resistivity measurements to provide a general characterisation of the physical properties.

## 2-Experimental

Polycrystalline samples with nominal composition La$_{0.8}$Cd$_{0.2}$MnO$_3$ and La$_{0.7}$Cd$_{0.3}$MnO$_3$ were prepared by solid-state reaction. Appropriate molar quantities of La$_2$O$_3$, CdO and MnO$_2$ were mixed, grounded and heat-treated at 800 $^0$C for 48 hours. After that, the samples were three times reground, pressed into pellets and sintered at 1000 $^0$C for 24 hours. We avoid using higher sintering temperatures to minimise the evaporation of Cd. Some samples from the same batch were also annealed at 800 $^0$C in a 1% $H_2$ and 99 % Ar atmosphere for 24 hours. We will call the samples in the following way: after annealing $La_{0.8}Cd_{0.2}MnO_3$ (S1), as-prepared $La_{0.8}Cd_{0.2}MnO_3$ (S2), after annealing $La_{0.7}Cd_{0.3}MnO_3$ (S3) and as-prepared $La_{0.7}Cd_{0.3}MnO_3$ (S4). Iodometric titration (see table II) showed that the amount of $Mn^{4+}$ ions increase from 15 % in sample S1 to 24 % in sample S4. $Mn^{4+}$ ions in $La_{1-x}Cd_xMnO_{3+\delta}$ are due not only to the partial substitution of La by Cd, but also to the excess of oxygen atoms $\delta$. We found, using a Scanning Electron Microscope, that the mean grain size (L) were between 0.58 and 1.05 $\mu$m.

Powder XRD patterns have been collected on a standard diffractometer using Cu K$\alpha$ radiation. The Raman scattering measurement were recorded in back scattering geometry by using a Renishaw RM 1000 system equipped with a DMLM series Leica microscope with a spectral resolution of approximately 2 cm$^{-1}$. The Ar$^+$ laser emission at 488 nm was used as excitation source. The magnetization measurements were performed using an MPMS SQUID magnetometer by Quantum Design at 4.2-400 K and in magnetic fields up to 15 kOe. Transport measurements were performed by the conventional dc-current four-probe method.

3- Results and Discussion

X-ray diffraction patterns reveal a perovskite phase and small amount of impurities of CdO in each sample. A Rietveld refinement of all the data, using a perovskite structure with a double unit cell, allowed us to study the preferential occupation of La or Cd on La sites. We did not found any preferential occupation of La or Cd. The fitting also included a second phase of CdO as impurities. Table I shows the lattice parameters and unit cell volumes for the perovskite phase, as well as, the percentage of CdO in the samples. The percentage of CdO impurities decreases from sample S1 to S4. The calculated volumes confirm a report from Troyanchuk et al.[4], where the substitution of La by Cd (instead of Ca) produces larger unit cell[7].

Table I. Lattice parameters, unit cell volumes and % of CdO for the four studied samples.

|    | a (Å) | b (Å) | c (Å) | $V_{unit-cell}$ (Å$^3$) | CdO (%) |
|----|-------|-------|-------|-------------------------|---------|
| S1 | 5.458 | 5.514 | 3.899 | 58.07                   | 7.3     |
| S2 | 5.447 | 5.506 | 3.884 | 57.63                   | 5.7     |
| S3 | 5.460 | 5.519 | 3.892 | 58.01                   | 2.8     |
| S4 | 5.449 | 5.498 | 3.875 | 57.52                   | 2.0     |

Table II. Data of composition, dimension, magnetic and resistivity for the four studied samples.

|    | $Mn^{4+}$ (%) | L ($\mu$m) | $\mu_S$ ($\mu_B$/Mn) | $T_C$ (K) | $\rho_{20K}$ ($\Omega$ cm) |
|----|---------------|------------|----------------------|-----------|----------------------------|
| S1 | 15            | 0.58       | 3.5                  | 165       | 1.9 x 10$^7$               |
| S2 | 19            | 0.91       | 3.5                  | 175       | 9.9 x 10$^3$               |
| S3 | 21            | 1.05       | 3.5                  | 165       | 3.1 x 10$^3$               |
| S4 | 24            | 1.05       | 3.4                  | 150       | 8.6 x 10$^2$               |

Figure 1 shows the Raman spectra at room temperature for the four studied samples. These spectra were reproducible for different points of each sample. The observed spectra are quite complex and reflect four different contributions. First, it could be seen for all samples three phonon bands, which are centered near 240, 432 and 614 cm$^{-1}$. These frequencies have been detected for different compositions of $A_{1-x}A'_xMnO_3$ and are consistent with the rear-earth vibrational mode, the Mn-O bending mode, and the Mn-O stretching mode of the $MnO_6$ octahedral, respectively[8-11]. Second, there is a broad multi-phonon like mode, between 200 and 650 cm$^{-1}$, probably induced by the presence of polarons. Third, using the high frequency part of the spectra, we found a "collision-dominated" electronic Raman scattering contribution associated with diffusive hopping of the carriers[8]. Finally, we noticed a frequency independent background, well visible at high frequencies. To our knowledge, this is the first Raman report for manganese perovskites doped with Cd.

Figure 2 shows the magnetisation versus temperature measured in a magnetic field of 1 T for all samples. Curves represent the transition from a low temperature ferromagnetic to a high temperature paramagnetic state. Curie temperatures, taken as the inflexion point of these curves, and the saturation magnetisation ($\mu_s$), taken at 20 K, change between 150 and 175 K and between 3.4 and 3.5 $\mu_B$/Mn (table II). Troyanchuk et al.[4,5] reported a Curie temperature of 150 K in a $La_{0.7}Cd_{0.3}MnO_3$ sample. However, the same paramagnetic to ferromagnetic transition between 150 and 160 K was also obtained by C. Ritter et al.[12] in $LaMnO_{3+\delta}$ compounds for values of $\delta$ between 0.07 and 0.15. These results make difficult to associate the magnetic phase transition only to the substitution of La by Cd or the excess of O atoms. On the other hand, Tc values for compounds of $La_{1-x}Ca_xMnO_3$ change between 150 and 250 when x goes from 0 to 0.3[13] demonstrating the ferromagnetic interaction is stronger in manganese perovskites doped with Ca. Saturation magnetisation values could be compared to a simplified theoretical model[14]. Taking the orbital momentum to be quenched in both $Mn^{3+}$ and $Mn^{4+}$, the magnetic moment reduces to the spin contribution g S $\mu_B$, where S is the spin of the ion (3/2 for $Mn^{4+}$ and 2 for $Mn^{3+}$) and g is the gyromagnetic factor (approximately 2 in both cases). For (1-x) $Mn^{3+}$ ions and x $Mn^{4+}$ ions, it becomes (4-x) $\mu_B$. In our case, using the results of x from the Iodometric titration, $\mu_s$ decreases from 3.85 to 3.76 $\mu_B$/Mn. The experimentally measured lower values could be due to the presence of impurities in the samples.

Figure 3 shows the resistivity in a logarithmic scale as a function of temperature for the four compounds studied. The resistivity increases between 300 and 20 K from 6 to 3 order of magnitude. Samples S1 and S2 show a monotonous increase in resistivity until the lowest measured temperature. Sample S3 also shows a monotonous grows with decreasing temperatures, but the first signs of an insulator metal transition start to appear. Meanwhile, sample S4 shows a maximum at about 82 K and a minimum near 29 K, before increasing again the resistivity with even lower temperatures. Troyanchuk et al.[4,5] have reported an insulator type increase of the resistivity with decreasing temperatures for $La_{1-x}Cd_xMnO_3$ with x between 0.25 and 0.5. However, the absolute value of the resistivity in our samples is from 1 to 2 order of magnitude lower at 77 K.

Resistivity values at 20 K are given in table II. It could be seen that an increase in the percentage of $Mn^{4+}$ ions correlates to a decrease in the resistivity at low temperatures and the gradual change in the shape of the curves. A transition to a metallic ferromagnetic ground state when the number of $Mn^{4+}$ increases is typical of manganese perovskites doped with Ca. However, the insulator metal transition in sample S4 is shifted to temperatures lower than the Curie temperature and the resistivity still have an insulator like increase at the lowest measured temperatures. In our opinion, although the general trends seem to be the same for both Ca and Cd doped manganese perovskites, the use of Cd ions weaken the double exchange interaction, that is believed to be partially responsible of the ferromagnetic metallic ground state. Troyanchuk et al.[4, 5] interpreted their results in Cd doped samples in the framework of a superexchange interaction.

The inset in figure 3 shows a plot of the same resistivities as a function of $(1/T)^{1/4}$. Above 200 K, all curves could be approximately fitted to a variable range hopping model. We also tried to fit the data to a thermally activated model with less success. Resistivity measurements in polycrystalline samples should be interpreted as a superposition of contributions from the grains and grain boundaries. Therefore, thin film or single crystal measurements are clearly necessary to further elucidate the above points.

4-Conclusion

We reported a general characterisation of the physical properties of four Cd doped manganese perovskites using X-rays diffraction, Iodometric titration, Raman, Magnetic and Electrical Resistivity measurements. Results are qualitatively similar to Ca doped samples.

Acknowledgment

J. López thanks the Condensed Matter Section and the Training in Italian Laboratories Program of the Abdus Salam ICTP for the economic support, which allowed him to do this work. We also thank the help of Franco Corticelli and Roberto Barboni.

Figure Captions

Fig. 1. Raman spectrum excited with a 2.54 eV laser for the four studied samples. Curves have been shifted vertically for clarity. The data is interpreted as a superposition of phonon and electronic contributions.

Fig. 2. Magnetisation versus temperature curves for the four studied samples. The magnetisation is represented in units of Bohr magneton per manganese ion. A paramagnetic to ferromagnetic phase transition is seen around 165 K.

Fig. 3. Resistivity curves in logarithmic scale versus temperature. The solid horizontal line indicates the Curie temperature interval found from magnetic measurements. A metal insulator transition is seen at 82 K for the sample with the higher number of $Mn^{4+}$ ions (S4). The inset shows the same curves of resistivity plotted as a function of $(1/T)^{1/4}$. For a clear visualization, resistivity of sample S2, S3 and S4 were multiplied by 5, 10 and 30, respectively.

Figure 1

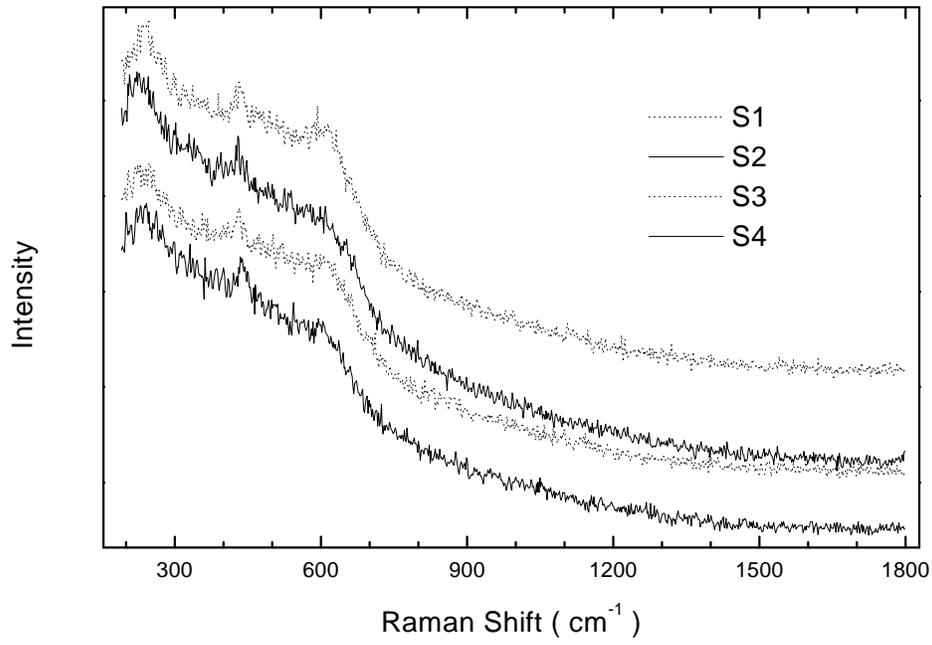

Figure 2

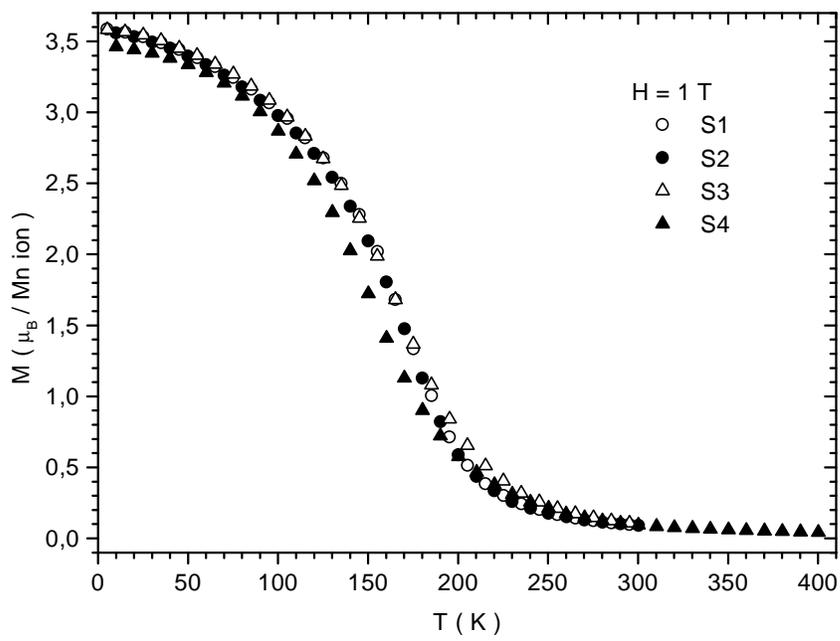

Figure 3

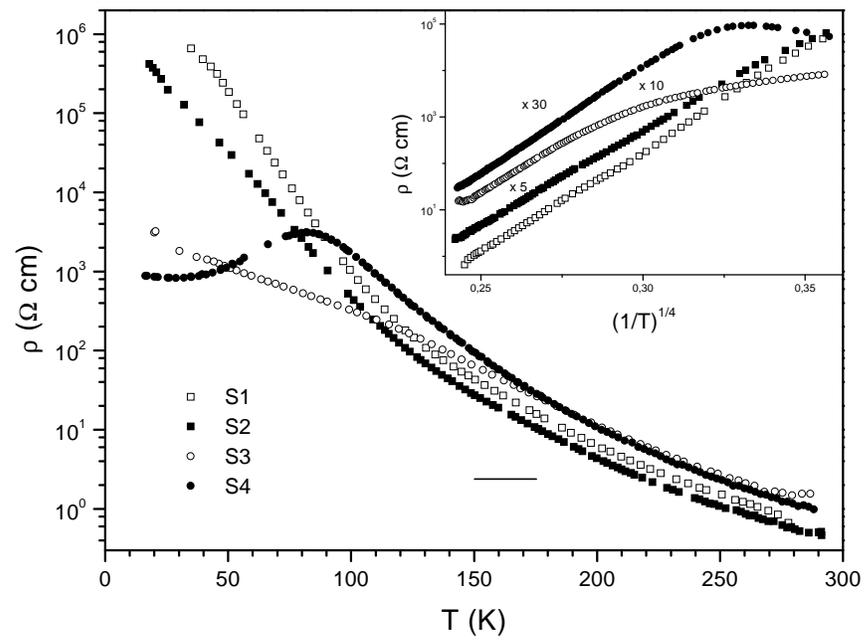